\documentclass[a4paper,twoside]{article}

\usepackage{epsfig}
\usepackage{epstopdf}
\usepackage{subcaption}
\usepackage{calc}
\usepackage{amssymb}
\usepackage{amstext}
\usepackage{amsmath}
\usepackage{amsthm}
\usepackage{multicol}
\usepackage{pslatex}
\usepackage{apalike}
\usepackage{algorithm2e}
\usepackage[bottom]{footmisc}
\usepackage{url}
\usepackage{xcolor}
\usepackage{enumitem}
\usepackage[most]{tcolorbox}
\usepackage{balance}

\definecolor{main}{HTML}{808080} 
\definecolor{sub}{HTML}{f2f2f2}  

\tcbset{
  sharp corners,
  colback=white,
  before skip=0.2cm,
  after skip=0.5cm
}

\newtcolorbox{findingsBox}{
  colback=sub,
  colframe=main,
  boxrule=0pt,
  leftrule=6pt,
  top=2mm,
  bottom=2mm,
  left=2mm,
  right=2mm
}

\usepackage{SCITEPRESS}

\begin{document}

\title{Chatbot-Based Assessment of Code Understanding in Automated Programming Assessment Systems}

\author{\authorname{Eduard Frankford\sup{1}\orcidAuthor{0009-0005-5959-4936}, Erik Cikalleshi\sup{1}\orcidAuthor{0009-0007-4349-4249} and Ruth Breu\sup{1}\orcidAuthor{0000-0001-7093-4341}}
\affiliation{\sup{1}University of Innsbruck, Department of Computer Science, Innsbruck, Austria}
\email{eduard.frankford@uibk.ac.at, erik.cikalleshi@student.uibk.ac.at, ruth.breu@uibk.ac.at}
}

\abstract{Large Language Models (LLMs) challenge conventional automated programming assessment because students can now produce functionally correct code without demonstrating corresponding understanding. This paper makes two contributions. First, it reports a saturation-based scoping review of conversational assessment approaches in programming education. The review identifies three dominant architectural families: rule-based or template-driven systems, LLM-based systems, and hybrid systems. Across the literature, conversational agents appear promising for scalable feedback and deeper probing of code understanding, but important limitations remain around hallucinations, over-reliance, privacy, integrity, and deployment constraints. Second, the paper synthesizes these findings into a Hybrid Socratic Framework for integrating conversational verification into Automated Programming Assessment Systems (APASs). The framework combines deterministic code analysis with a dual-agent conversational layer, knowledge tracking, scaffolded questioning, and guardrails that tie prompts to runtime facts. The paper also discusses practical safeguards against LLM-generated explanations, including proctored deployment modes, randomized trace questions, stepwise reasoning tied to concrete execution states, and local-model deployment options for privacy-sensitive settings. Rather than replacing conventional testing, the framework is intended as a complementary layer for verifying whether students understand the code they submit.}

\onecolumn \maketitle \normalsize \setcounter{footnote}{0} \vfill

\section{Introduction}
Large Language Models have fundamentally changed programming education and assessment. The growing complexity of modern coding skills has made AI a necessary tool for supporting the learning process \cite{manorat_artificial_2025}. However, this shift also introduces substantial assessment risks. While tools such as ChatGPT can provide personalized feedback, they can also generate complete solutions that students may submit without understanding \cite{lin_athena_2025}.

This creates a difficult situation for assessment. At present, most Automated Programming Assessment Systems (APASs) still rely primarily on unit tests and basic static checks to validate functional correctness. These mechanisms remain useful, but they are weak proxies for conceptual understanding when students can generate plausible code with LLMs \cite{vintila_avert_2024}. Consequently, assessment must move beyond the final artifact and examine whether students can explain the logic, execution, and design decisions behind their submissions \cite{lehtinen_automated_2023}.

In this paper, the term \textit{conversational agent} is used as an umbrella term for chatbots and dialog-based tutoring or assessment systems that maintain multi-turn natural-language interaction with a learner and adapt subsequent turns based on previous answers \cite{debets_chatbots_2025}. Conversational agents offer a promising way to bridge the assessment gap. Instead of relying only on static quizzes or black-box functional tests, these systems can engage students in dialogue and use scaffolding techniques to probe how well they understand the code they submitted \cite{cheng_integrating_2025}. Isolated examples of this idea already exist, such as systems that ask questions about a learner's code \cite{lehtinen_automated_2023} or tools designed to verify authorship and understanding \cite{vintila_avert_2024}. However, there is still no widely adopted design synthesis for integrating these conversational elements into APASs in a transparent and scalable manner.

Accordingly, this paper is structured in two parts. First, it presents a systematic review of chatbot-based and conversational assessment approaches in programming education. Second, it derives a framework proposal from the resulting findings. The overall goal is to identify effective practices, categorize technical approaches, surface limitations, and synthesize design guidance for a next generation of APASs that can assess both code functionality and student understanding.

\section{Related Work}
The use of AI in computer science classrooms is expanding rapidly. A recent review covered more than 100 papers to categorize how such tools are being used, from course design to grading \cite{manorat_artificial_2025}. While that review highlights the utility of AI for feedback generation and grading support, it addresses the educational landscape broadly. The present study narrows the focus to conversational agents used to verify knowledge during programming assessment.

When looking at chatbots specifically, Debets et al. found that most existing tools are designed for teaching rather than testing \cite{debets_chatbots_2025}. In their review of 71 papers, they noted that many chatbots rely on platforms such as Dialogflow but often lack a strong theoretical foundation. The present work focuses on the intersection of chatbots, program comprehension, and formal assessment.

Traditional assessment systems usually rely on unit tests and static code analysis. However, passing a test case does not necessarily mean that a student understands the underlying concept. Research by Lehtinen et al. shows that students often achieve \textit{unproductive success}, where they pass the assignment through trial and error or copying without truly understanding the code \cite{lehtinen_automated_2023,canico_integrating_2025}. Approaches such as asking students to explain their own code have been proposed to address this gap \cite{lehtinen_automated_2023}, but a systematic account of how such conversational verification can be integrated into an automated grading pipeline is still missing.

\section{Methodology}
\subsection{Review Questions and Design Objective}
This paper intentionally separates review questions from the framework proposal. The systematic review answers two research questions, and the framework section then synthesizes the findings into a design proposal.

\begin{itemize}
    \item \textbf{RQ1:} How can conversational assessment approaches in programming education be categorized based on their technical implementation, pedagogical strategies, and assessment methods?
    \item \textbf{RQ2:} What are the benefits, limitations, and pedagogical implications of chatbot-based assessment approaches with respect to scalability, fairness, and learning outcomes?
\end{itemize}

Based on the answers to RQ1 and RQ2, the second part of the paper derives a \textbf{design objective}: to propose a framework for integrating a conversational agent into an APAS in a way that is transparent, grounded in program analysis, and scalable enough for larger cohorts.

\subsection{Search Strategy and PRISMA-Guided Review Protocol}
\begin{figure}[t]
    \centering
    \includegraphics[width=1\linewidth]{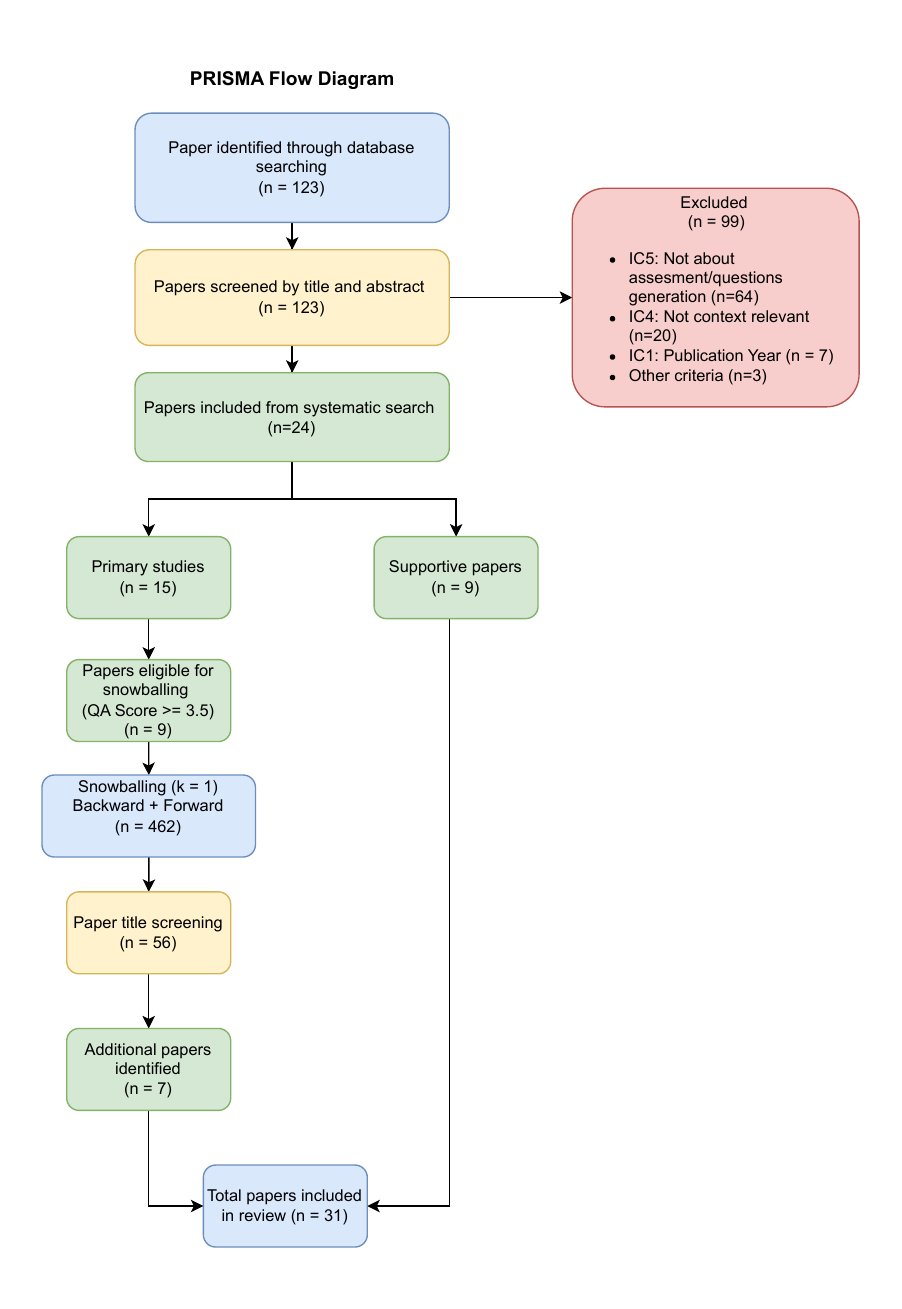}
    \caption{PRISMA-style flow diagram of identification, screening, eligibility, and inclusion.}
    \label{fig:prisma}
\end{figure}

The review followed a PRISMA-style workflow covering identification, screening, eligibility, and inclusion. Figure~\ref{fig:prisma} summarizes the selection process. To improve reproducibility, the full search strings, quality assessment rubric, and row-level screening decisions are provided in the supplementary materials on Zenodo.\footnote{\url{https://zenodo.org/records/18335209}}

\subsubsection{Information Sources}
Google Scholar was used as the primary meta-search engine to retrieve literature and citation links across the ACM Digital Library, IEEE Xplore, ScienceDirect, arXiv, and SpringerLink. To reduce the risk of missing papers because of ranking artifacts, targeted validation searches were also performed directly on these sources where appropriate. Searches were conducted between 1 October 2025 and 30 November 2025.

\subsubsection{Search Strings}
Five search strings were developed using the PICOC framework (Population, Intervention, Comparison, Outcome, Context) to capture different aspects of conversational assessment in programming education \cite{kitchenham}. The keyword groups were defined as follows:

\begin{itemize}
    \item Population: students in computer science or programming courses; APASs
    \item Intervention: chatbots, conversational agents, conversational assessment, intelligent tutoring systems
    \item Comparison: traditional methods, manual review, non-conversational assessment
    \item Outcome: code understanding, learning outcomes, explanation quality, authorship verification
    \item Context: university-level education in the LLM era
\end{itemize}

\begin{figure*}[t]
\centering
\includegraphics[width=0.6\textwidth]{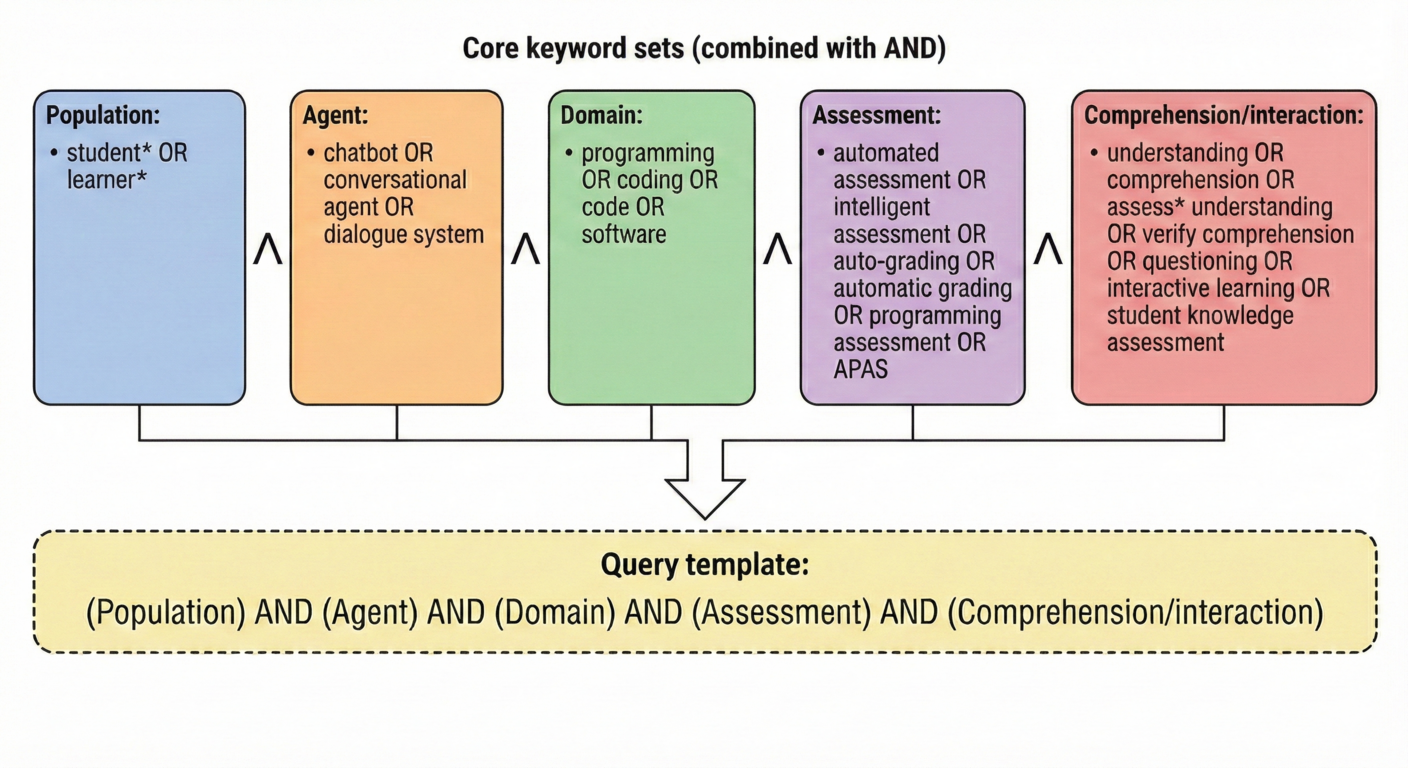}
\caption{Keyword groups and query template used to derive the five search string variants.}
\label{fig:search-strings}
\end{figure*}

To keep the main text concise, Figure~\ref{fig:search-strings} summarizes the keyword groups and query template. The five full search strings are provided in the supplementary materials on Zenodo.

\subsubsection{Search Execution and Screening Process}
Given the rapid evolution of LLMs in education, a \textit{saturation-based scoping review} methodology was adopted. This approach was selected because the domain remains small, fast-moving, and heterogeneous.

\begin{enumerate}
\item \textbf{Initial Search:} Each search string was executed across the selected sources between October and November 2025. Results were sorted by relevance where possible. Title and abstract screening were combined into a single stage because titles alone were often insufficient to determine relevance.

\item \textbf{Iterative Screening:} Search results were reviewed page by page. Papers were retained for full-text review when they appeared to address conversational assessment, chatbot-based questioning, code understanding, or intelligent tutoring in programming education.

\item \textbf{Saturation-Based Stopping:} For each search string and source combination, screening continued until 3 consecutive result pages yielded no new relevant papers. The saturation point was usually reached after pages 3 to 5 of Google Scholar results.

\item \textbf{Duplicate Removal and Decision Logging:} Duplicates were removed manually during screening. A screening log captured the source, search string identifier, retrieval window, screening stage, inclusion decision, and exclusion reason for each full-text decision.
\end{enumerate}

\subsubsection{Snowballing}
To ensure comprehensive coverage, one iteration ($k=1$) of snowballing was conducted \cite{Wohlin}, applying both backward and forward snowballing to the papers that passed the initial inclusion and exclusion criteria. Forward snowballing examined papers that cited the included studies using Google Scholar's \textit{Cited by} feature, while backward snowballing reviewed the reference lists of included papers.

During this process, the same inclusion and exclusion criteria were applied. Papers already identified during the systematic search were excluded to avoid duplication. Very short papers such as workshop abstracts and posters were also excluded. Many cited papers were initially assessed by title and publication year. A substantial share of the retrieved work focused on tutoring, debugging assistance, or one-way code explanations rather than using conversational interaction to verify student understanding during assessment.

\subsubsection{Selection Criteria}
To ensure relevance and quality, studies were selected according to the following criteria. Inclusion was limited to peer-reviewed papers or reputable gray literature published between 2018, marking the post-transformer era \cite{attention}, and 2025. Selected studies were required to be written in English and to address programming or computer science education, including transferable methods from closely related domains such as mathematics or logic. Crucially, the primary focus had to be on assessing code understanding via conversational agents, chatbots, or intelligent tutoring systems, or explicitly discussing the impact of LLMs on programming assessment.

Records were excluded if they were promotional material, advertisements, or opinion pieces without technical or empirical substance. Studies focused on non-programming domains without a clear link to computer science education, or studies that mentioned chatbots or assessment only tangentially, were discarded. Research published prior to 2018 and articles whose full text was inaccessible through institutional access or author contact were also excluded.

\section{Review Findings}
\subsection{Categorization of Approaches}
This section answers RQ1. Based on the systematic analysis of the primary studies, conversational assessment approaches in programming education can be grouped into three dominant categories: (1) rule-based or template-driven systems, (2) LLM-based systems, and (3) hybrid systems. Using the dominant architecture of the concrete assessment systems discussed in the corpus, hybrid approaches appeared most frequently (5/12), followed by rule-based or template-driven systems (4/12) and LLM-based systems (3/12).

\subsubsection{Rule-Based and Template-Driven Systems}
Rule-based approaches represent the most structured method for conversational assessment. These systems rely on deterministic algorithms and predefined templates to generate questions. Typically, they use static and dynamic code analysis to understand the student's code \cite{alshaikh_experiments_2021,santos_jask_2022,thomas_stochastic_2019}. Static analysis often employs Abstract Syntax Trees (ASTs) to identify code constructs such as variable declarations, loops, and conditional statements. Dynamic analysis simulates the execution of the code to track variable values and runtime behavior.

Question generation is deterministic, meaning that the same input consistently produces the same questions. This predictability supports consistency and auditability, but it also limits flexibility. Technically, these systems parse code to extract structural elements such as variable declarations, loop constructs, conditional statements, and function definitions. Questions are then generated by matching these elements against template libraries \cite{santos_jask_2022,stankov_smart_2023,thomas_stochastic_2019}.

For instance, when a \texttt{for} loop is detected, the system may generate questions about initialization, continuation conditions, termination, and incrementation. Some rule-based systems extend this approach with dynamic execution and use simulated runs to generate questions about runtime behavior, variable values, and execution traces \cite{santos_jask_2022,thomas_stochastic_2019}.

The primary advantage of rule-based systems lies in their consistency, modest computational overhead, scalability, and immediate feedback \cite{thomas_stochastic_2019,santos_jask_2022}. However, they face significant limitations regarding flexibility and variety. They cannot easily handle code patterns outside their predefined templates, which can lead to repetitive questioning \cite{alshaikh_experiments_2021}. Building larger and more expressive template libraries also demands substantial time and maintenance effort.

\subsubsection{LLM-Based Systems}
LLM-based systems use models such as GPT, Llama, or Mistral to generate natural and contextually relevant questions without relying on predefined templates \cite{kargupta_instruct_2024,wang_how_2024}. Unlike rule-based approaches, these systems can sustain multi-turn dialogues that adapt to student responses and maintain conversational context. However, they face a central challenge: general-purpose LLMs are optimized to be helpful assistants and may therefore provide direct solutions instead of guiding students through Socratic questioning \cite{kargupta_instruct_2024}.

To address this, more advanced systems implement structured workflows that constrain LLM behavior toward pedagogically sound interaction. These systems draw on established teaching methods such as the Socratic method and scaffolding theory to shape how questions are generated and sequenced \cite{alshaikh_experiments_2021,al-hossami_socratic_2023}. One notable idea is the \textit{trace-and-verify} workflow. TreeInstruct \cite{kargupta_instruct_2024}, for example, tracks specific concepts or bugs as binary state variables and uses this state to build dynamic question trees, where sibling questions probe the same misconception from different angles.

The main strengths of LLM-based systems are their ability to generate natural-sounding questions, handle diverse code patterns, and adapt pedagogical strategies. Their weaknesses include inconsistent outputs, higher computational cost, and dependence on the capabilities and reliability of the underlying model \cite{al-hossami_socratic_2023,kargupta_instruct_2024,wang_how_2024}.

\subsubsection{Hybrid Systems}
Hybrid systems strategically combine the structure of rule-based approaches with the flexibility of machine-learning components. This architecture acknowledges that coherent conversation flow and reliable fact extraction benefit from explicit structure, while tasks such as response evaluation, paraphrasing, and distractor generation benefit from the pattern-recognition capacity of ML \cite{al-hossami_socratic_2023,vimalaksha_digen_2021,wang_training_2025}.

ChatDAC \cite{chuang_chatgpt-based_2025} exemplifies this approach by integrating dynamic assessment into a chatbot. It uses GPT-4 to evaluate student explanations against a reference reason and compute a similarity score. This score determines the level of scaffolding provided. A low score triggers broad hints, while a higher score allows more focused guidance. Similarly, Sakshm AI \cite{gupta_sakshm_2025} uses a chatbot named Disha within explicit Socratic guardrails and provides context-aware hints and structured feedback without revealing direct code solutions. AVERT \cite{vintila_avert_2024} further illustrates the value of combining deterministic program evidence with conversational verification when authorship and understanding must both be examined.

The primary advantage of hybrid systems is that they can balance coverage with control. They reduce the fragility of pure rule-based systems and the unpredictability of pure LLM systems, but they also introduce higher architectural complexity and additional integration effort \cite{al-hossami_socratic_2023,vimalaksha_digen_2021}.

\begin{findingsBox}
\textbf{RQ1 Summary: Categorization of Approaches}
\begin{itemize}
  \item \textbf{Rule-based or template-driven:} These systems rely on deterministic AST analysis and predefined templates to ensure consistency and immediate feedback \cite{santos_jask_2022,thomas_stochastic_2019,stankov_smart_2023}. Their main limitation is rigidity and long-term maintenance cost.
  \item \textbf{LLM-based:} These approaches use generative models to facilitate natural, multi-turn Socratic dialogue that adapts to student input \cite{kargupta_instruct_2024,wang_how_2024,lin_athena_2025}. Their key risks are hallucinations, direct answer-giving, and inconsistent grading behavior.
  \item \textbf{Hybrid:} These frameworks combine deterministic code analysis with generative dialogue management \cite{chuang_chatgpt-based_2025,gupta_sakshm_2025,vintila_avert_2024}. They appear most suitable when scalability, transparency, and pedagogical control are all required.
\end{itemize}
\end{findingsBox}

\subsection{Benefits, Limitations, and Pedagogical Implications}
This section answers RQ2. The integration of chatbot-based assessment systems into programming education offers clear advantages for scalability and personalized learning, but it also introduces technical, behavioral, and pedagogical challenges.

\subsubsection{Benefits: Scalability, Engagement, and Self-Improvement}
The primary benefit of automated conversational assessment is the ability to provide immediate and personalized feedback at a scale that human instructors cannot easily match. Empirical studies consistently report measurable improvements in student performance. For instance, the Socratic Author system demonstrated a 43\% learning gain in programming knowledge compared to a control group \cite{alshaikh_experiments_2021}. Similarly, the implementation of ChatDAC resulted in a significant increase in post-test scores, and stronger engagement with tiered hints correlated positively with final exam performance \cite{chuang_chatgpt-based_2025}.

Beyond academic performance, these systems also affect student psychology and engagement. Research on PythonPal suggests that personalized chatbot feedback can reduce transactional distance in online learning and foster a stronger sense of engagement \cite{palahan_pythonpal_2025}. Furthermore, the AIvaluate system suggests that such tools can reduce teacher burden in performance-based assessments while maintaining assessment quality in larger cohorts \cite{yusuf_towards_2025}.

\subsubsection{Limitations: Reliability, Over-Reliance, and Integrity}
Despite their potential, conversational assessment tools face major limitations. A central technical challenge is the tendency of LLMs to hallucinate, omit relevant execution details, or produce inconsistent evaluations. Research using the Let's Ask AI framework showed that even strong models such as GPT-4 can still make errors similar to novice programmers, for example by misreading execution paths \cite{lehtinen_lets_2024}.

Behavioral challenges are equally important. Students may \textit{game} the system or become over-reliant on AI support. An exploratory study by Rahe et al. \cite{rahe_how_2025} observed repeated debugging loops in which students excessively prompted the chatbot for fixes rather than trying to understand the underlying logic. This over-reliance threatens academic integrity, particularly when students can produce correct code or polished explanations without mastery \cite{elhambakhsh_evaluating_2025}. Students have also reported disappointment with generic or repetitive responses from AI tutors when those systems fail to adapt to the actual context of the submission \cite{frankford_ai-tutoring_2024}.

\subsubsection{Pedagogical Implications: The Gap Between Writing and Understanding}
The most important pedagogical implication in this literature is the visibility of \textit{unproductive success}, where students produce functionally correct code without understanding how it works. Findings from the Jask system show a strong contrast between performance on static code-structure questions and dynamic execution questions: while students achieved success rates above 80\% on static-structure questions, performance dropped below 50\% on dynamic execution questions \cite{santos_jask_2022}. This gap is concerning because code tracing skill is a strong predictor of overall programming competence \cite{lehtinen_automated_2023,stankov_smart_2023}.

Interaction analyses further suggest that the most effective systems encourage reflection rather than merely providing answers \cite{khor_exploring_2025}. Taken together, these findings support a shift away from purely functional grading and toward conversational validation that verifies conceptual depth and reasoning quality \cite{vintila_avert_2024}.

\begin{findingsBox}
\textbf{RQ2 Summary: Benefits, Limitations, and Implications}
\begin{itemize}[leftmargin=1em, itemsep=-1mm]
  \item \textbf{Benefits:} Conversational agents can provide personalized feedback at scale, improve learning gains, and reduce teacher burden in large cohorts \cite{alshaikh_experiments_2021,yusuf_towards_2025}.
  \item \textbf{Limitations:} Technical risks such as hallucinations \cite{lehtinen_lets_2024} are compounded by behavioral risks such as gaming, over-reliance, and dependence on AI-generated fixes \cite{rahe_how_2025}.
  \item \textbf{Pedagogical implications:} The literature repeatedly exposes a gap between producing correct code and understanding it \cite{santos_jask_2022}. This supports shifting assessment from purely functional testing toward conversational verification of reasoning.
\end{itemize}
\end{findingsBox}

\subsection{Synthesis of Design Requirements}
The review findings suggest five requirements for a practical conversational layer in APASs: (1) grounding questions in deterministic code facts, (2) supporting multi-turn probing instead of one-shot explanation, (3) discouraging answer copying and LLM-generated explanations through runtime-specific prompts, (4) preserving privacy and transparency in data handling, and (5) separating formative support from high-stakes summative use. These requirements motivate the framework proposed in the next section.

\section{Framework Proposal Derived from the Review}
Synthesizing the findings from RQ1 and RQ2, this section proposes a Hybrid Socratic Framework for integrating conversational verification into APASs. The proposal is intended as a design synthesis rather than a validated end product. Its purpose is to balance the conversational fluency of LLMs with the reliability of deterministic code analysis.

\begin{figure*}[t]
    \centering
    \makebox[\textwidth][c]{\includegraphics[width=\textwidth]{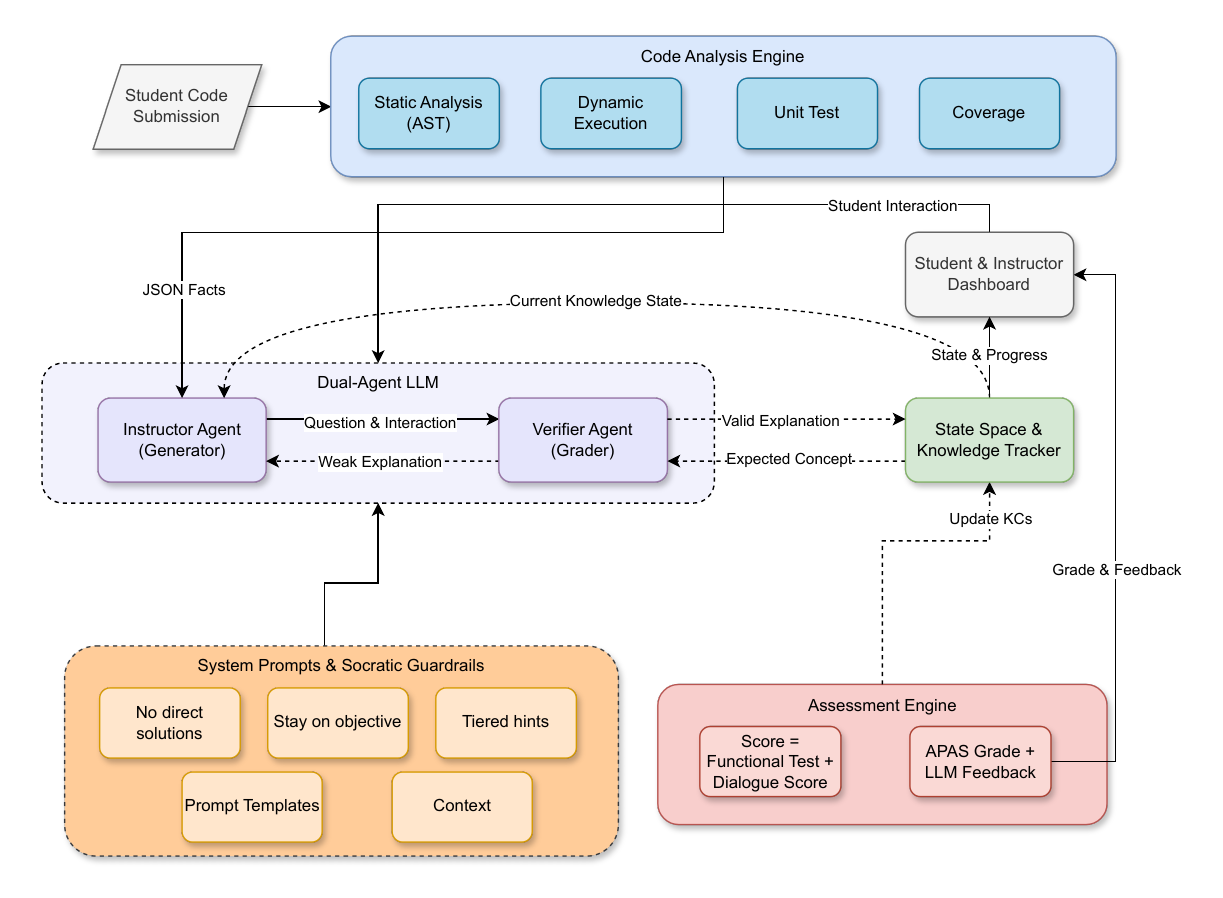}}
    \caption{Hybrid Socratic Framework for chatbot-based assessment in APASs.}
    \label{fig:hybrid-socratic-framework}
\end{figure*}

The framework uses a hybrid architecture in which the conversational layer is constrained by static and dynamic analysis of the submitted code. The Socratic method serves as the pedagogical foundation because it requires students to articulate intermediate reasoning rather than merely supply an answer. The overall architecture is shown in Figure~\ref{fig:hybrid-socratic-framework}.

\subsection{Essential Components}
The proposed framework consists of five modules operating in a closed loop:

\begin{itemize}
    \item \textbf{Code Analysis Engine:} This component uses static analysis and dynamic execution to extract deterministic facts about the student's submission. \\
    \textit{Example:} If a loop never terminates, the engine can identify the relevant line and the missing state transition, providing ground truth that constrains later prompting.

    \item \textbf{State Space and Knowledge Tracker:} This module maintains a persistent model of the student's understanding by tracking knowledge components and misconceptions across turns. \\
    \textit{Example:} If a student can explain variable scope but fails to explain pointer arithmetic, the tracker prioritizes follow-up questions about pointer behavior.

    \item \textbf{Dual-Agent Conversational Core:} Splitting the generative task into two specialized agents separates questioning from evaluation:
    \begin{enumerate}
        \item \textbf{Instructor Agent:} Formulates Socratic questions from the extracted code facts.\\
        \textit{Example:} \textit{Look at variable \(i\) in the loop body. What value does it have immediately before the final iteration executes?}
        \item \textbf{Verifier Agent:} Evaluates student responses against a reference reason grounded in the code-analysis output. \\
        \textit{Example:} It compares the student's explanation against the trace-based reference explanation and estimates whether mastery has been demonstrated.
    \end{enumerate}

    \item \textbf{Assessment Engine:} A hybrid grading layer combines evidence from code functionality and dialogue quality. \\
    \textit{Example:} Code that passes tests but is paired with a weak or inconsistent explanation triggers additional questioning and may lower the final score.

    \item \textbf{Socratic Guardrails:} These constraints prevent the agents from reverting to a solution-giving assistant role. \\
    \textit{Example:} If a student asks for the fix directly, the system must redirect the interaction toward a trace, boundary condition, or design rationale instead of supplying code.
\end{itemize}

\subsection{Integrity Safeguards Against LLM-Generated Explanations}
A central challenge is that students may also use LLMs to generate explanations. For this reason, the framework should not rely on generic \textit{Why does your code work?} questions. Instead, it should bind explanation requests to concrete and potentially randomized execution evidence.

\begin{enumerate}
    \item \textbf{Proctored deployment for high-stakes use:} In high-stakes settings, the conversational layer should be deployed in supervised labs, controlled viva-style sessions, or remote proctoring contexts. Unproctored use is more suitable for formative practice.

    \item \textbf{Randomized trace questions:} The Code Analysis Engine should generate trace questions from runtime states that are specific to the submitted program and, where possible, to randomized inputs. This makes generic LLM-generated answers less useful because the student must refer to concrete execution states.

    \item \textbf{Stepwise reasoning tied to execution states:} Rather than accepting a single polished explanation, the system should require students to reason through successive states, for example by predicting the next value of a variable, identifying the last valid array access, or explaining why a loop terminates.

    \item \textbf{Adaptive follow-up questions:} When the Verifier Agent detects vague or generic language, the Instructor Agent should issue a follow-up that narrows the reasoning space, such as changing the input, asking for the next trace state, or focusing on a specific branch.
\end{enumerate}

Together, these safeguards reduce the usefulness of copy-pasted explanations and shift the burden of proof toward real-time reasoning about the student's actual code.

\subsection{Agent Prompting Strategy}
To ensure that the LLMs behave within the Socratic guardrails, each agent requires a dedicated system prompt. These prompts are adapted from design patterns reported in Sakshm AI, TreeInstruct, and ChatDAC \cite{gupta_sakshm_2025,kargupta_instruct_2024,chuang_chatgpt-based_2025}. Rather than relying on a single generic instruction, the framework separates question generation from response evaluation and constrains both tasks with explicit roles, inputs, and output expectations.

\subsubsection{Instructor Agent Prompt}
This agent receives the code-analysis output, the current dialogue state, and the target concept to be probed. In practical terms, the prompt defines the agent as a Socratic tutor whose task is to ask short, focused questions tied to concrete program facts, such as a specific variable update, branch condition, loop boundary, or runtime state. It is explicitly instructed not to reveal the fix, not to provide corrected code, and not to confirm correctness too early. Instead, it should move from simpler comprehension checks toward deeper reasoning prompts and use the dialogue history to adapt the next question to the student's last answer. The overall purpose of the prompt is therefore to turn deterministic evidence from the code-analysis layer into guided questioning that triggers explanation rather than solution copying.

\subsubsection{Verifier Agent Prompt}
This agent evaluates student responses against a reference reason grounded in the facts produced by the Code Analysis Engine. Its role is not to define the truth space from scratch, but to compare the student's explanation against deterministic program evidence and then decide whether additional scaffolding is needed. The prompt therefore frames the agent as a constrained evaluator that checks conceptual alignment with the expected execution logic, tolerates variation in wording, and returns a structured judgment about whether the answer is sufficient, partially correct, or incorrect. It also determines which misconception or missing step should be targeted next. In this way, the Verifier Agent supports consistent assessment while keeping the grading process anchored in the actual behavior of the submitted program rather than in an unconstrained model interpretation.

\subsection{Design Principles}
The framework is governed by three core design principles.

\subsubsection{Two-Tier Assessment Strategy}
To reduce lucky guessing and superficial interaction, the framework adopts a two-tier approach similar to ChatDAC.
\begin{itemize}
    \item \textit{Tier 1 (Selection):} The chatbot presents a scenario or code-behavior query generated by the Instructor Agent.
    \item \textit{Tier 2 (Explanation):} The student must provide a natural-language explanation for the selected answer.
\end{itemize}
Scores are calculated by the Verifier Agent using a weighted formula:
$$ Score = 20 + (Similarity \times 0.8) $$
The base score of 20 acknowledges a correct selection in Tier 1, but high scores are only achievable with a valid explanation in Tier 2.

\subsubsection{Socratic Guardrails}
To prevent the LLM from reverting to an assistant role and giving direct solutions, strict prompting guardrails are enforced. The system redirects off-topic questions back to the current learning objective and refuses to generate code until the student has demonstrated conceptual understanding through dialogue.

\subsubsection{Scaffolded Decomposition}
Complex programs are decomposed into smaller logic units. Instead of assessing an entire program in one step, the system uses code analysis to ask targeted questions about specific branches, states, or logic blocks, so that the student must demonstrate understanding component by component.

\subsection{Proof of Concept Implementation and Current Limitations}
To validate the technical feasibility of the proposed framework, a functional prototype was developed in a Python-based stack. Source code and a demo video are available on Zenodo.\footnote{\url{https://zenodo.org/records/18335209}}

The prototype uses \textbf{Streamlit} for the interactive student interface and session state management. While the initial version used \textbf{Gemini 2.0 Flash}, the updated prototype is \textbf{model-agnostic}. It supports multiple backends, including hosted inference APIs and local transformer models through \texttt{llama-cpp-python}, which improves flexibility and enables privacy-sensitive deployment. A fallback mode allows operation without active API keys by using pre-generated datasets.

\begin{figure}[t]
    \centering
    \includegraphics[width=1\linewidth]{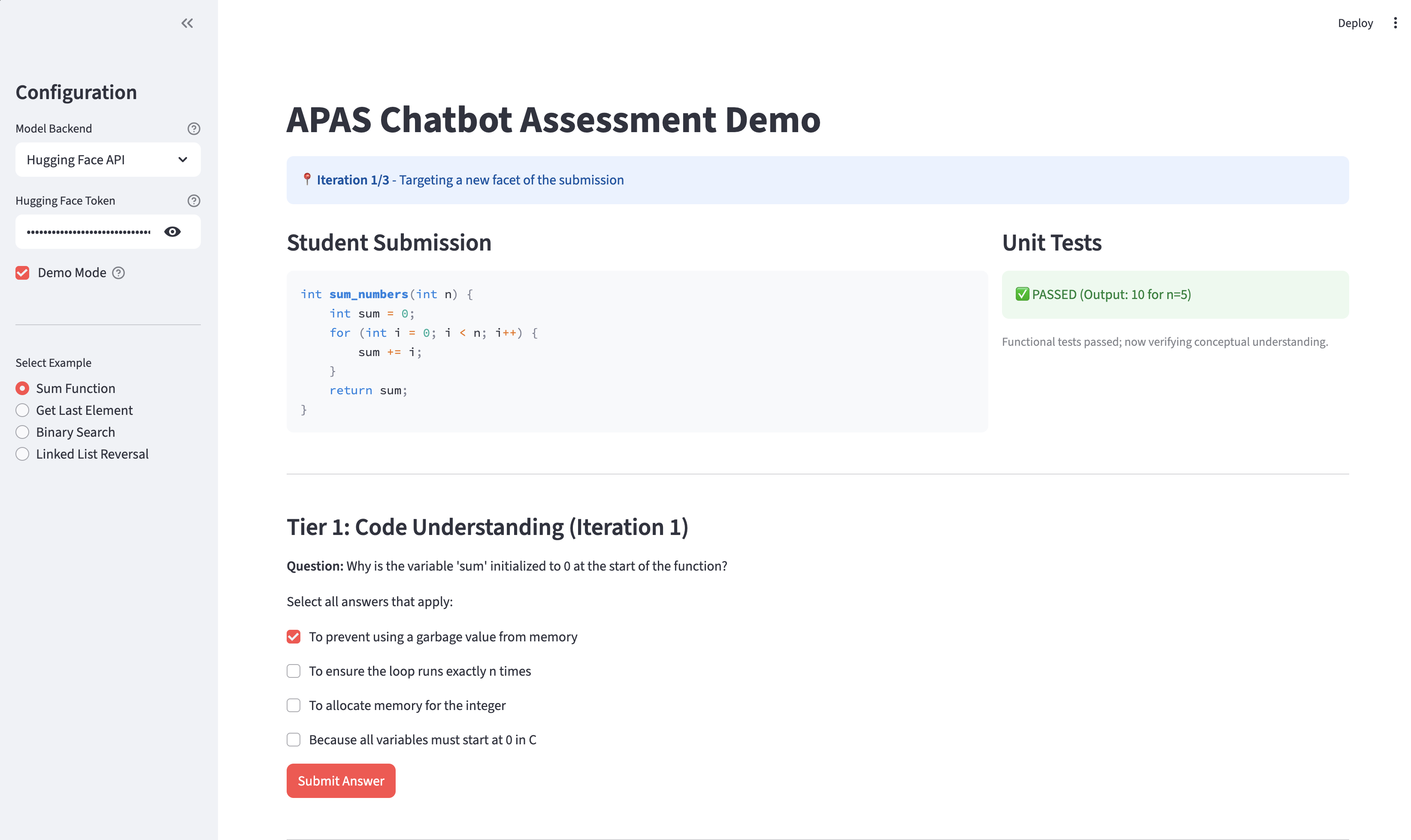}
    \caption{Proof of concept, Tier 1: scenario-based multiple-choice question generated by the Instructor Agent.}
    \label{fig:poc-tier1-mcqs}
\end{figure}

\begin{figure}[t]
    \centering
    \includegraphics[width=1\linewidth]{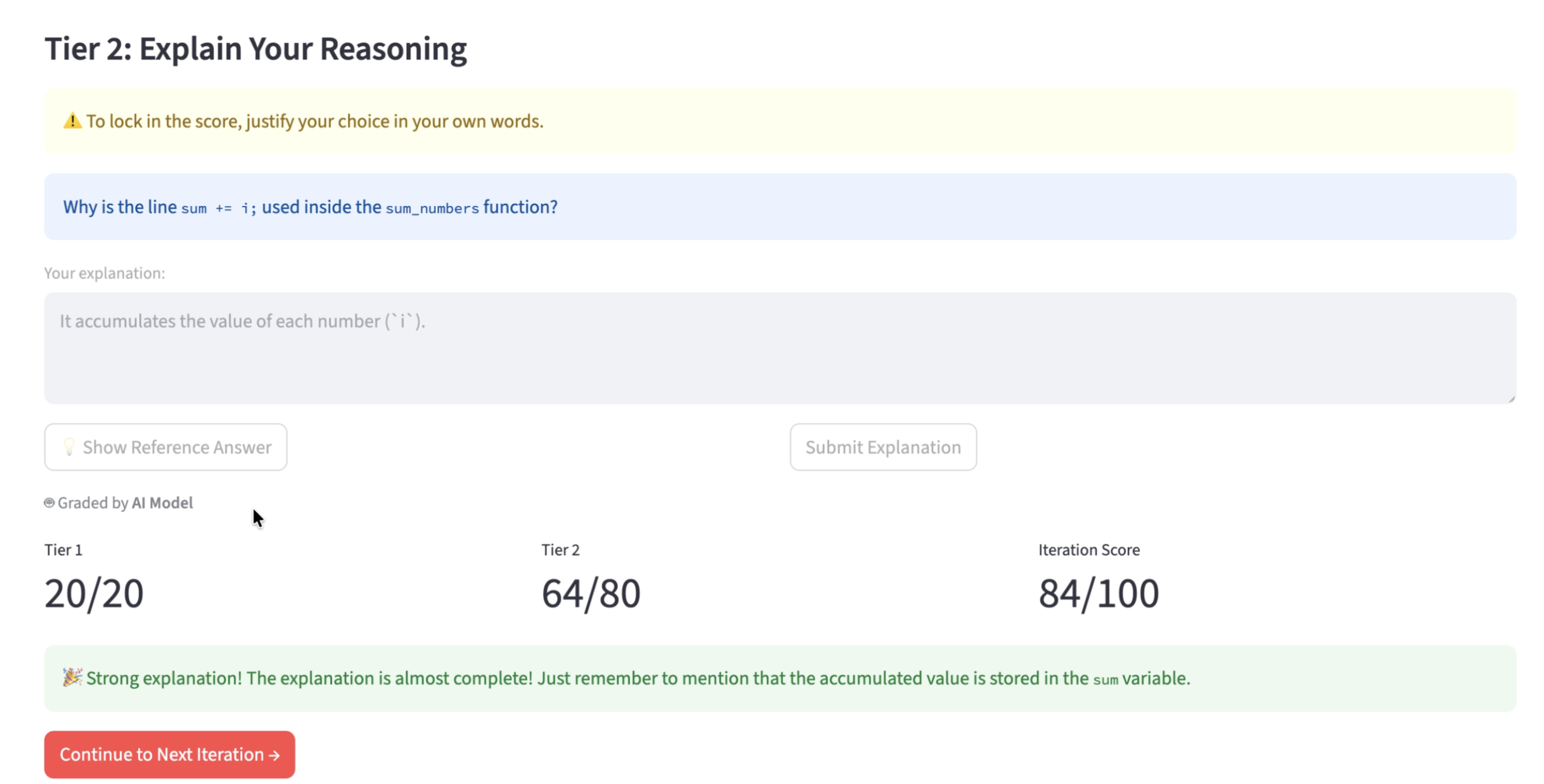}
    \caption{Proof of concept, Tier 2: open-ended explanation graded by the Verifier Agent.}
    \label{fig:poc-tier2-explanation}
\end{figure}

The implementation enforces the dual-agent architecture illustrated in Figures~\ref{fig:poc-tier1-mcqs} and \ref{fig:poc-tier2-explanation}:

\begin{enumerate}
    \item \textbf{Instructor Agent:} Receives raw C code input and generates scenario-based multiple-choice questions targeting specific conceptual elements such as pointer management.
    \item \textbf{Verifier Agent:} Acts as a semantic grader by comparing the student's natural-language explanation against a hidden reference reason generated from the same grounded context.
\end{enumerate}

The current prototype nevertheless has important limitations. It demonstrates the end-to-end conversational flow, but it does not yet constitute a production-ready APAS component. In particular, deterministic unit-testing and regression-testing hooks are not yet integrated into a complete assessment pipeline, language support is limited, no classroom deployment study has been conducted, and no inter-rater reliability study has compared automated judgments with human examiners. In addition, latency, operational cost, model drift, and adversarial prompt resistance have not yet been evaluated systematically. For these reasons, the prototype should currently be interpreted as a feasibility demonstration rather than validated evidence of assessment reliability.

\section{Discussion}
The review and the framework proposal together indicate that APASs are moving from a primary focus on functional outputs toward stronger verification of learners' reasoning. Dialogue-based verification increases the effort required to submit AI-generated code without understanding, particularly when questions are tied to execution traces and code-specific states.

A second implication is the growing importance of cognitive depth in automated questioning. Rather than posing generic prompts, effective conversational assessors align questions with explicit comprehension targets such as code tracing, control-flow reasoning, or explanation of design tradeoffs. This suggests that future APASs should map chatbot turns to well-defined comprehension targets rather than treating conversation as an unstructured add-on.

A third implication is the persistent tradeoff between validity and scalability. LLM-based systems scale and support natural multi-turn interaction but introduce reliability risks through hallucinations and inconsistent outputs, while rule-based systems offer stronger auditability but weaker coverage and adaptability. The literature therefore points toward hybrid architectures that ground conversational assessment in deterministic program analysis.

\subsection{Technology Acceptance and Classroom Integration}
Technology acceptance is likely to determine whether such systems are used successfully in practice. Students need to understand whether the chatbot is acting as a tutor, an examiner, or both. Instructors need clear controls over prompt policies, scoring thresholds, appeals, and when the conversational layer is formative versus summative. Transparent communication about what is being assessed and how explanations are evaluated is therefore essential for trust and adoption.

\subsection{Ethics, Fairness, and Accountability}
Ethical concerns arise when conversational systems influence grades. Students must be protected against opaque grading logic, biased language judgments, and inconsistent treatment across language backgrounds or communication styles. A grounded hybrid architecture helps by tying evaluation to program facts, but this does not remove the need for human oversight, appeal procedures, and ongoing validation against instructor judgments.

\subsection{Digital Sustainability}
Digital sustainability also deserves attention. Vendor-dependent cloud deployments may be difficult to maintain over time because of changing APIs, pricing, and model availability. Local-model options improve institutional control and privacy, but they introduce hardware, maintenance, and energy costs. Sustainable adoption therefore depends not only on pedagogical effectiveness but also on maintainable infrastructure, reproducible prompt and model versioning, and realistic operating costs.

\section{Threats to Validity}
While this study followed a systematic methodology, several risks to validity remain.

\subsection{Technological Volatility}
A significant challenge in reviewing LLM-based tools is the pace of technological change. Several studies in the corpus rely on models or frameworks that were already being replaced during the review period. Some reported limitations may therefore reflect the maturity of specific models rather than fundamental limits of conversational assessment as a paradigm.

\subsection{Framework Design Validity}
The proposed Hybrid Socratic Framework is derived from the review, but it has not yet been validated through a full classroom study. As a result, design choices such as similarity thresholds, scaffolding sequences, or the weighting of explanation quality remain provisional. The framework should therefore be interpreted as a reasoned synthesis, not as a validated assessment standard.

\section{Conclusion and Future Work}
Large Language Models make purely functional programming assessment increasingly insufficient because students can now generate correct code without necessarily understanding it. This paper therefore addressed conversational assessment in two stages.

First, it reviewed the emerging literature on conversational agents in programming assessment and grouped the identified approaches into rule-based or template-driven, LLM-based, and hybrid systems. The review indicates that conversational agents can provide scalable and personalized probing of code understanding, but it also highlights substantial risks related to hallucinations, over-reliance, privacy, fairness, and academic integrity.

Second, based on these findings, the paper proposed a Hybrid Socratic Framework that combines deterministic code analysis with a constrained conversational layer. The framework is intended to support code-understanding verification by grounding questions in program facts, requiring stepwise reasoning, and separating question generation from explanation evaluation.

Overall, the literature and the proposed framework suggest that the future of programming assessment will likely involve conversational verification as a complement to conventional testing rather than as a replacement for it. The most promising path is not to ban AI outright, but to use carefully constrained conversational agents to examine whether students can reason about the code they submit.

\subsection{Future Work}
Several directions remain open. First, future work should validate the framework empirically through classroom deployments and comparisons with instructor-led oral checks. Second, stronger knowledge-tracing mechanisms could help the system maintain a more stable model of student understanding across multiple sessions. Third, multimodal extensions may eventually allow students to explain code through traces, diagrams, or memory-state sketches rather than text alone. Finally, future studies should investigate when conversational assessment is appropriate for formative use, when it can support summative decisions, and how deployment choices affect privacy, fairness, and student behavior.

\balance

\end{document}